\documentclass[fleqn,twoside,twocolumn,nofootinbib,showkeys]{revtex4} 
\usepackage[sec,doi]{ujp_UTF8} 

\begin{document}
\title[UHE hadronic physics at Auger]
{Ultra-high-energy hadronic physics at the Pierre Auger Observatory: muon measurements}%
\author{J.~Ebr}
\affiliation{FZU -- Institute of Physics of the Czech Academy of Sciences}
\address{Na Slovance 1999/2, Prague, Czech Republic}
\email{ebr@fzu.cz}
\author{ for the Pierre Auger Collaboration}
\affiliation{Observatorio Pierre Auger, full author list:  http://www.auger.org/archive/authors2024\_10.html}
\address{Av. San Mart{\'i}n Norte 304, 5613 Malarg\"{u}e, Argentina}

\udk{№ УДК/UDC} \razd{\secix}

\autorcol{J.~Ebr for the Pierre Auger Collaboration}%

\setcounter{page}{1}%

\begin{abstract}
The Pierre Auger Observatory, the world’s largest observatory of ultra-high-energy cosmic rays (UHECR), offers a unique insight into the properties of hadronic interactions occurring in air showers at energies well above those reached at human-made accelerators. The key probe into the hadronic interactions has, for a long time, been the number of muons arriving at the ground, which can be directly measured at Auger for energies up to 10 EeV using dedicated underground muon detectors or estimated through the observation of highly inclined showers using the surface detector of the Observatory. Further information can be obtained using the hybrid character of the Observatory, which allows the simultaneous observation of the longitudinal development of the shower with the fluorescence (and lately also radio) detector and the ground signal with the surface detector. Several different analyses using hybrid data show a discrepancy between the predictions of simulations based on the latest hadronic interaction models and data. This discrepancy has been long interpreted as a deficit in the number of muons predicted by the simulations with respect to the data. A new analysis using a global fit of the data on selected hybrid showers has shown that the disagreement between models and data is more complex and also involves the predictions for the depths of the maxima of the longitudinal shower development. At the same time, measurements of shower-to-shower fluctuations using inclined hybrid events show good agreement with the predictions, suggesting that the observed muon discrepancy is rather the result of a gradual accumulation of small changes during the shower development than of a major change in the properties of the first interaction. Recently, the Observatory has undergone an upgrade, which includes several components aimed at a significant improvement in the measurement of the muon content of the air showers. 
\end{abstract}


\maketitle

\section{Introduction}

Ultra-high-energy cosmic rays (UHECR) allow us to study hadronic interactions at energies far above the capabilities of current accelerators. The energies of the most energetic cosmic rays observed are above $10^{20}$ eV -- if such a particle (usually assumed to be a proton or a heavier nucleus) interacts with a nucleon in a nucleus in the air which is stationary, the center-of-mass energy exceeds 400 TeV. However due to their rarity (less than one particle per km$^2$ per century), the UHECR interactions can be only observed through the observations of the extensive air showers created by those particles in the atmosphere. At ultra-high energies, the showers consist of many generations of interactions and, at their maximal development, can contain billions of particles. As these particles spread over many square kilometers (of a location that cannot be predicted beforehand), only a tiny fraction of them for each shower can be detected with reasonably-sized ground detectors. To complement this information, the development of the shower in the atmosphere can be observed from a distance by the fluorescence light -- and recently also the radio waves -- created during the passage of charged particles through the atmosphere. 

Even in the largest UHECR detectors, the amount of information obtainable for each shower is fairly limited. From the fluorescence profile, mainly the integral (proportional to the energy of the primary cosmic ray) and the depth of the shower maximum are extracted. The ground detectors do not identify individual particles and their momenta and provide instead only the density of all charged particles as a function of the distance from the shower axis, with possible additional information available through the time structure of the signal. Separate densities for muons can be obtained if additional types of detectors are installed. 

However as it turns out, even this limited information is sensitive to the properties of the hadronic interactions at ultra-high energies. This can be seen already using a highly simplified model of the shower cascade \cite{matthews} (known as the Heitler-Matthews model) where each hadronic interaction in the shower produces all three species of pions in equal numbers and with the energy equally shared between them and is characterized by its charged multiplicity $N_\mathrm{ch}$ and elasticity $\kappa$. The neutral pions immediately decay into two photons, each of which initiates and electromagnetic (EM) cascade where pair production and bremsstrahlung alternate in producing electron, positrons and more photons. If the hadronic cascade stops when the pion energy reaches the critical energy $\xi_\mathrm{c}^\pi\approx 20\,\mathrm{GeV}$, $\xi_\mathrm{c}^\mathrm{e}$ is the critical energy for the EM cascade, $\lambda_\mathrm{r}$ is the radiation length in air and $X_0$ is the depth of the first interaction (as given by the relevant cross-section) then for a shower intitiated by a particle of energy $E_0$ and nucleon number $A$, we can estimate the depth of shower maximum $X_\mathrm{max}$ (in terms of amount of mass traversed in $g/cm^2$) and the number of muons at ground $N_\mu$ as 
\begin{equation}
X_\mathrm{max}\approx\lambda_\mathrm{r}\ln(E_0/\xi_\mathrm{c}^\mathrm{e})+X_0-\lambda_\mathrm{r}(\ln(3N_\mathrm{ch})+\ln A)
\end{equation}
\begin{multline}
N_\mu\approx\left(\frac{E_0}{\xi_\mathrm{c}^\pi}\right)^\beta A^{(1-\beta)}\\
\beta\approx 1-\frac{\kappa}{3\ln N_\mathrm{ch}}  > 0.9.
\end{multline}
In this approximation, both quantities depend on  the interaction properties - $X_\mathrm{max}$ depends on the interaction cross-section (through $X_0$) and multiplicity, while $N_\mu$ depends on multiplicity and elasticity. Using more detailed Monte Carlo simulations, it can be shown that the dependence is more complicated \cite{ulrich}, but the main features of the model are conserved.

\begin{figure}
\vskip1mm
\includegraphics[width=\column]{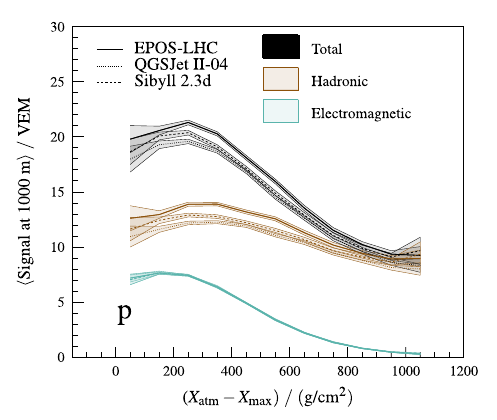}
\vskip-3mm\caption{Ground signal at 1000 meters from shower axis as a function of distance to the shower maximum for primary protons separated into electromagnetic and hadronic contributions, for different hadronic interaction models. From \cite{vicha}.  }
\label{fig1}
\end{figure}

An interesting corollary of the Heitler-Matthews model is that with each successive generation of hadronic interactions, more energy is transferred to the electromagnetic part of the cascade. The fraction of EM energy is given as
\begin{equation}
\frac{E_\mathrm{em}}{E_0}=1-\left(\frac{E_0}{\xi_\mathrm{c}^\pi A}\right)^{\beta-1}    
\end{equation}
As the number of generations increases with primary energy, so does this ratio. For a primary proton at $10^{19}$ eV, about 90 \% of primary energy ends up in electrons, positrons and photons. However, while the EM particles carry most of the energy, their ground signal is mostly predicted by the depth of the maximum $X_\mathrm{max}$. As a function of the distance to $X_\mathrm{max}$, the EM signal depends only very weakly on the choice of the hadronic interaction model (Fig.~\ref{fig1}).  This phenomenon is known as \textit{shower universality} and for its  precise description, the particles in the shower must be separated into four components, three of which (muons, EM particles from muon decays and EM particles created locally in hadronic jets) can be understood as the \textit{hadronic} component of the shower, while the remaining part is the \textit{pure} electromagnetic component \cite{ave}. As the hadronic component is at larger distances from the shower axis mostly dominated by the muonic part, measurements that quantify either can often be directly compared. 

\begin{figure}
\vskip1mm
\includegraphics[width=\column]{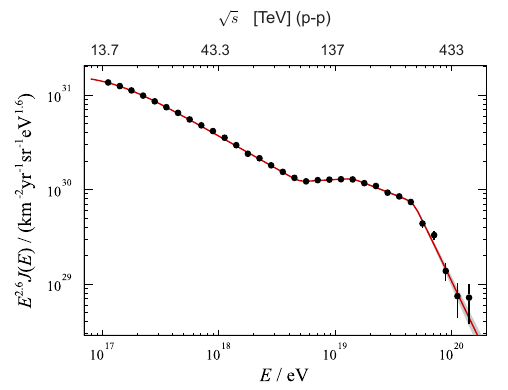}
\vskip-3mm\caption{UHECR spectrum with lab energies translated into nucleon-nucleon center-of-mass energies, data from \cite{spectrum}.  }
\label{fig2}
\end{figure}

As both the depth of maximum and number of muons depend on both the interaction properties and the primary mass $A$ (both a priori unknown), they cannot be directly interpreted separately. However combining the two variables can already lead to strong statements about hadronic interactions, if a consistent choice of the mass composition of the primary beam cannot be made to describe both measurements simultaneously. As UHECR exceed the LHC energies in the center-of-mass frame already at $10^{17}$ eV, there is a large span of energies where the UHECR observations are the main experimental input for the understanding of hadronic interactions, chiefly through muon measurements (Fig.~\ref{fig2}). The Pierre Auger Observatory \cite{auger}, located near Malarg\"{u}e, Argentina, is the world's largest UHECR detector, consisting primarily of a surface detector array spanning 3000 km$^2$ and 27 fluorescence telescopes overlooking the surface array. Thanks to its leading position in the field, the Observatory is also the largest source of insights into hadronic interactions at ultra-high energies. While the fluorescence detector directly measures the depth of shower maximum, the surface detector cannot straightforwardly separate the hadronic and electromagnetic components of the ground signal and thus various novel methods have been devised for hadronic interaction studies.

\section{Direct muon measurements}

Muons can be separated simply by putting a sufficient amount of mass between the detector and the air shower, so that most other particles are absorbed before reaching the detector. At Auger, this is  achieved in two different ways, either by observing showers at very large zenith angles (where the mass is the large amount of atmosphere to be traversed before reaching ground) or by putting additional muon detectors underground. 

\begin{figure}
\vskip1mm
\includegraphics[width=\column]{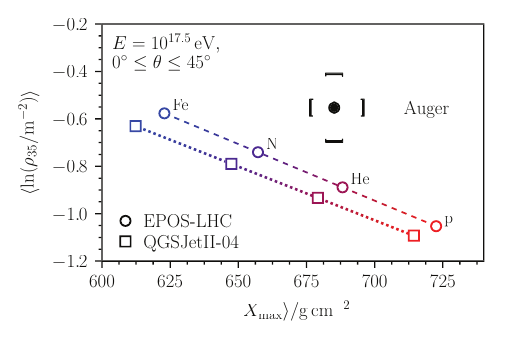}
\vskip-3mm\caption{Muon number from the engineering array of the Underground Muon Detector and mean depth of shower maximum, compared with predictions of hadronic interaction models for different primary particles, from \cite{underground}.  }
\label{fig3}
\end{figure}

The Underground Muon Detector has been deployed over many years in a specific area of the Auger array, where the surface detectors are more densely spaced to increase the sensitivity to lower-energy showers, with a target date for full deployment in 2025 \cite{UMD}. The data from an earlier engineering array of seven surface detector stations equipped with muon detectors has been already published \cite{underground}. Each of the muon detectors consists of four plastics scintillators buried 2.3 meters under ground so that the energy cutoff for vertical muon is roughly 1 GeV. As the muon number strongly correlates with primary energy (see Eq.~(2)), an independent measurement of the primary energy is needed for any muon measurement. In the case of the UMD, the primary energy is independently measured by the nearby surface detectors (based on an energy estimator calibrated using the fluorescence measurements). 

Due to the small surface area covered by the engineering array, the measurement can be carried out only for relatively low energies between $2\times10^{17}$ and $2\times10^{18}$ eV. The resulting muon number is compatible with pure iron (as the heaviest nucleus expected in astrophysical cosmic rays) across the entire energy range, however such an interpretation is at odds with the measured depth of shower maxima which suggest a much lighter composition, as it can be seen in Fig.~\ref{fig3}. Thus, even at energies immediately above that of the LHC, we already see a possible discrepancy between data and hadronic interaction models. This can be related to the fact that even at LHC energies, the predictions of the models may differ as the LHC detectors do not cover the entire kinematic range of produced particles. In particular they lack coverage in regions of high pseudorapidity which turn out to be the most important for the development of air showers due to the kinematics in the laboratory frame.

\begin{figure}
\vskip1mm
\includegraphics[width=\column]{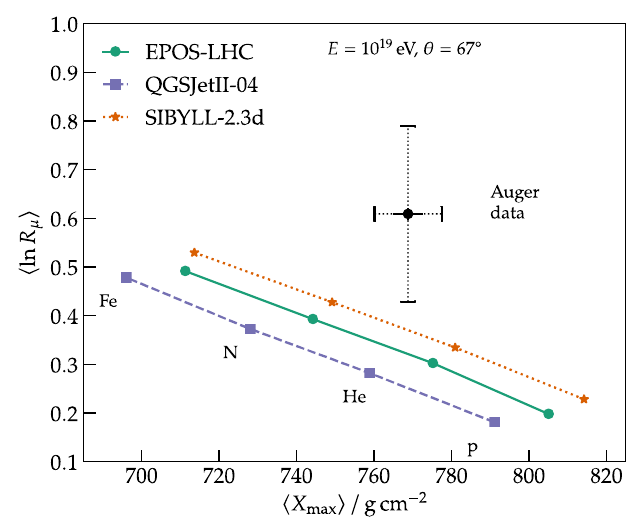}
\vskip-3mm\caption{Muon number from inclined hybrid events and mean depth of shower maximum, compared with predictions of hadronic interaction models for different primary particles, from \cite{fluct}.  }
\label{fig4}
\end{figure}

Using inclined showers, we can almost smoothly continue the measurement to higher energies, starting at $4\times10^{18}$ eV. Since in the case of inclined showers the muons are detected directly by the surface detectors, another independent measurement of the primary energy is needed. In the hybrid analysis \cite{inclined,fluct} this is naturally provided by the fluorescence detector for showers observed by both detectors -- however this corresponds only to roughly 13 \% of all showers observed by the surface detector since the operation of the fluorescence detector is limited by daylight, moonlight and weather. Still, 281 high-quality events were recorded with zenith angles in the range between 62 and 80 degrees where muons dominate the ground signal. For each shower, the signal footprint on the ground (the shape of which is strongly affected by the geomagnetic field and attenuation due to the long paths of the muons in the asmosphere) was compared to a corresponding simulated muon map and a scaling factor expressing the relative moun number found. 

\begin{figure}
\vskip1mm
\includegraphics[width=\column]{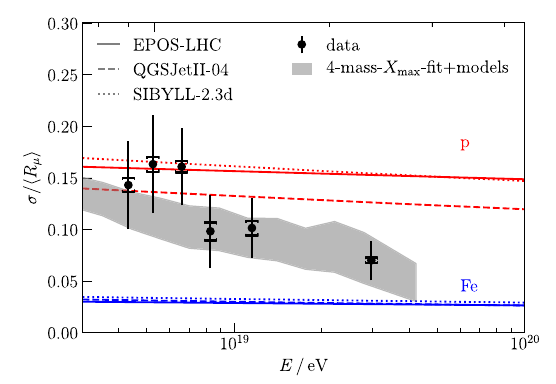}
\vskip-3mm\caption{Relative fluctuations of muon number compared with predictions of hadronic interaction models for proton and iron and for the mass composition model based on $X_\mathrm{max}$ measurements (gray band), from \cite{fluct}.  }
\label{fig5}
\end{figure}

In this case, the measured muon number increases with energy even above that predicted for iron primaries, even though it stays just barely compatible with iron within systematic uncertainty. Comparison with the primary composition inferred from the depths of shower maxima however again shows a significant discrepancy between the models and the data (Fig.~\ref{fig4}). These two measurements (using underground detectors and inclined showers) can be considered a clear illustration of the "muon problem" or "muon puzzle". When the data are interpreted using a primary composition consistent with the depths of shower maxima, the simulations using the current hadronic interaction models significantly underestimate the observed amount of muons. A similar trend can be seen also in data from other experiments, see \cite{whisp} for a combined analysis.

An interesting insight into the origin of the muon puzzle can be gleaned by measuring the shower-to-shower fluctuations of the number of muons in the hybrid inclined showers \cite{fluct}. Such a measurement requires careful study of the resolution in the measurement of both the muon number and energy so that the intrinsic fluctuations can be separated. Unlike the absolute muon numbers, the measured relative fluctuations (Fig.~\ref{fig5}) are well compatible with the predictions of the hadronic interaction models. We expect the fluctuations in the muon number at ground to be mostly affected by the fluctuations in the first interaction, as fluctuations in parallel interactions in subsequent generations have the tendency to balance each other out. The observation of a discrepancy in the overall number of muons but not in the relative fluctuations thus implies that we observe a cumulative effect over many generations of interactions, not an abrupt change in physics at the very highest energies.

\begin{figure}
\vskip1mm
\includegraphics[width=\column]{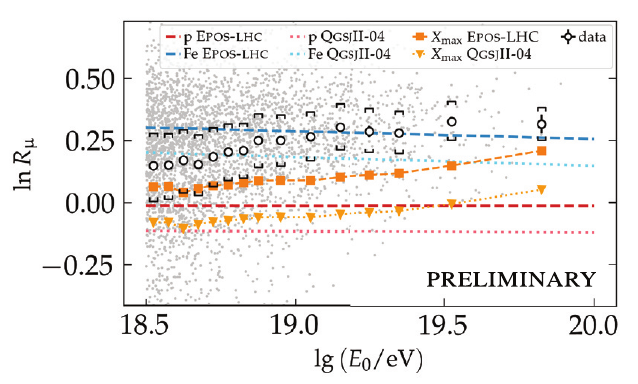}
\vskip-3mm\caption{Muon number in vertical hybrid showers determined using shower universality (open circles) compared to the muon number predicted by simulations when assuming a mass composition compatible with the relevant $X_\mathrm{max}$ distributions, from \cite{uni}.  }
\label{fig6}
\end{figure}

\section{Muons in vertical hybrid showers}

For "vertical" events (with zenith angle smaller than 60 degrees), the ground signal is a mixture of electromagnetic and muon signals. We can exploit shower universality to separate the two contributions either at the level of individual showers or their entire sets. If we assume that the electromagnetic signal as a function of distance to shower maximum is well described by the simulations, we can then attribute any discrepancy between the simulations and data to inadequate prediction of the number of muons or of the depth of shower maximum.

At the level of individual events, it is possible to use shower universality to construct a model of the surface signal as a function of distance from the shower axis for each shower based on its arrival geometry, primary energy, $X_\mathrm{max}$ and relative number of muons \cite{uni}. For hybrid showers, primary energy and  $X_\mathrm{max}$ are determined from fluorescence data and thus the number of muons is the only free parameter of the model and can be easily fitted. The mean observed number of muons in vertical hybrid showers is again larger than what would be expected for a set of events with primary masses corresponding to the observed $X_\mathrm{max}$ distributions at the given energy -- this discrepancy is roughly 15 \% for the hadronic model EPOS-LHC and 30 \% for QGSJETII-04 (Fig.~\ref{fig6}).

\begin{figure}
\vskip1mm
\includegraphics[width=\column]{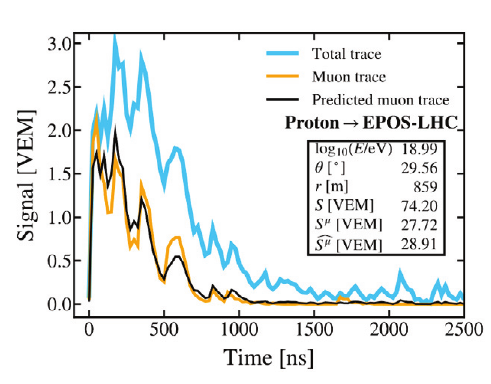}
\vskip-3mm\caption{An example of the prediction of the muon signal by a neural network (black) based on the overall signal in the surface detector station (blue), compared to the true simulated value (orange), from \cite{neural}.}
\label{fig7}
\end{figure}

\begin{figure*}
\vskip1mm
\includegraphics[width=\textwidth]{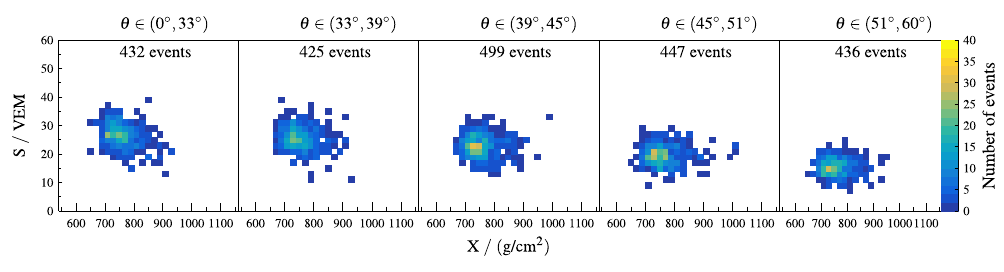}
\vskip-3mm\caption{Distributions of depths of maxima and ground signals for observed hybrid events between  $10^{18.5}$--$10^{19}$ eV, from \cite{vicha}.}
\label{fig8}
\end{figure*}

\begin{figure}
\vskip1mm
\includegraphics[width=\column]{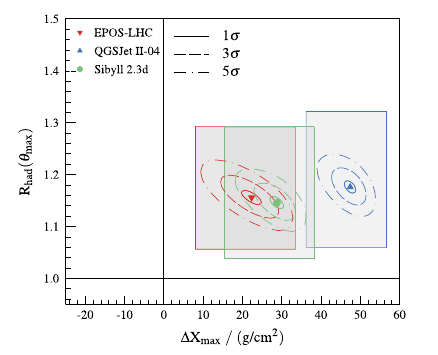}
\vskip-3mm\caption{The fitted values of the change in hadronic signal and shift in $X_\mathrm{max}$ needed for the best match between the simulations of a given hadronic interaction model and Auger data, for larger zenith angles around 55 degrees, from \cite{vicha}.}
\label{fig9}
\end{figure}

It is worth noting here that additional muon information is encoded in the time trace of the signal recorded in the surface detector and thus the amount of muons can in principle be estimated even on an station-by-station basis. Due to the complicated nature of the time traces, this has always been a difficult task, but recently encouraging results have been obtained using Recurrent Neural Networks. A study \cite{neural} based on simulations shows that the networks can be trained so that when given the time trace measured by the surface detector and geometric parameters of the shower, it can reliably predict what was the purely muon signal (Fig.~\ref{fig7}). This method has however not been yet applied to the data.

Instead of determining the muon content of each event, we can compare the observed and predicted distribution of ground signal to quantify any possible discrepancy. Because of the dependence of the predicted signal on primary composition and the complementary nature of $X_\mathrm{max}$ and $N_\mu$
with respect to hadronic interactions, it is natural to compare 2-dimensional distributions of $X_\mathrm{max}$ and ground signal, optimally taken at 1000 meters from shower axis for an array with the spacing of Auger (Fig.~\ref{fig8}). In the recent analysis \cite{vicha}, high-quality events between $10^{18.5}$ and $10^{19}$ eV were split into zenith-angle bins, adjusted to a reference energy and fitted with simulated templates of sets of p, He, O and Fe showers with free parameters being the fractions of the individual nuclei and a rescaling parameter $R_\mathrm{had}$ for the hadronic part of the signal. This hadronic part is simply defined as a fraction of the overall signal as determined by simulations, depending on zenith angle and primary mass. 

Interestingly, such a fit describes the Auger data relatively poorly. To achieve a better description, it is necessary to introduce an additional parameter, which is the shift of $X_\mathrm{max}$ with respect to the simulations. This is, for simplicity, implemented as a uniform shift of the $X_\mathrm{max}$ scale, independent on primary mass, and the change in ground signal induced by the shift is taken into account separately. This procedure was repeated for three different hadronic interaction models and $R_\mathrm{had}$ was allowed to vary with zenith angle. In all cases the fit consistently requires both a change in the hadronic signal between 15 to 25 \% and a shift of the $X_\mathrm{max}$ scale between 20 to 50 g/cm$2$ (Fig.~\ref{fig9}), with zero modifications excluded at the level of $5\sigma$ even when taking into account systematic uncertainties of the experiment. While independently such modifications of hadronic interactions are not dramatic, it can be shown that this particular combination is difficult to achieve simply by modifying the macroscopic characteristics of the interactions such as multiplicity, elasticity and cross-section \cite{mochi}.

Note that if the $X_\mathrm{max}$ shift indeed reflects physical reality, then the mass composition interpretation of Auger data changes and with it also changes the amount of discrepancy found in all of the other muon measurements presented above. Thus the consistency between the different methods to quantify the muon puzzle cannot be judged by a simple numerical comparison of the results.

\section{Outlook}

While the body of evidence for the need to adjust the hadronic interaction models in the light of the Auger data is now large, many details of the muon puzzle remain unclear. One of the ways to improve our understanding of UHECR extensive air showers is through the AugerPrime upgrade of the Observatory, the deployment of which is now being finalized \cite{prime}. The upgrade consists of many individual parts: scintillator-based surface detectors and radio detectors are added to each surface detector station to improve muon/EM separation for lower and higher zenith angles respectively. The electronics of the surface detector stations is being upgraded not only to cope with the additional data from the added detectors but also for better temporal resolution to improve the measurement of the time structure of the traces. A small photomultiplier is added to each station to improve the dynamical range. Also the area equiped with the Underground Muon Detectors, described in Sec.~2, is being expanded. All of these upgrades will dramatically improve the muon measurements described in this paper, possibly to an extent where we will be able to directly study the possible mechanisms for alleviating the muon puzzle.

\vskip3mm \textit{Acknowledgement.} This work was co-funded by the EU and supported by the Czech Ministry of Education, Youth and Sports through the project CZ.02.01.01/00/22\_008/0004632, GACR (Project No. 24-13049S) and CAS (LQ100102401).

\end{document}